\documentclass[conference]{IEEEtran}
\usepackage{color}
\usepackage[english]{babel}
\usepackage{graphicx}
\usepackage{epstopdf}
\usepackage{caption}
\usepackage{times}
\usepackage{multirow}
\usepackage[algo2e,linesnumbered]{algorithm2e} 
\usepackage{algorithmic}
\usepackage{algorithm}
\usepackage{psfrag}
\usepackage[shell]{gnuplottex}
\usepackage{mathenv}
\usepackage{array}
\usepackage[T1]{fontenc}
\usepackage{array}
\usepackage{mdwmath}
\usepackage{multicol}
\usepackage{dblfloatfix}
\usepackage{amsmath}
\usepackage{amssymb}
\usepackage{float}
\usepackage{url}
\makeatletter
\newcommand*{\rom}[1]{\expandafter\@slowromancap\romannumeral #1@}
\makeatother

\begin{document}

\title{Interference Avoidance Algorithm (IAA) for Multi-hop Wireless Body Area Network Communication}

\author{\IEEEauthorblockN{Mohamad Jaafar Ali\IEEEauthorrefmark{1}, Hassine Moungla\IEEEauthorrefmark{1}, Ahmed Mehaoua\IEEEauthorrefmark{1}}\IEEEauthorblockA{\IEEEauthorrefmark{1}LIPADE, University of Paris Descartes, Sorbonne Paris Cit\'{e}, 45 rue des saints p\`{e}res, 75006, Paris, France \\
Email: \{mohamad.ali; hassine.moungla; ahmed.mehaoua\}@parisdescartes.fr}}
  
\maketitle
\begin{abstract}
In this paper, we propose a distributed multi-hop interference avoidance algorithm, namely, IAA to avoid co-channel interference inside a wireless body area network (WBAN). Our proposal adopts carrier sense multiple access with collision avoidance (CSMA/CA) between sources and relays and a flexible time division multiple access (FTDMA) between relays and coordinator. The proposed scheme enables low interfering nodes to transmit their messages using base channel. Depending on suitable situations, high interfering nodes double their contention windows (CW) and probably use switched orthogonal channel. Simulation results show that proposed scheme has far better minimum SINR (12dB improvement) and longer energy lifetime than other schemes (power control and opportunistic relaying). Additionally, we validate our proposal in a theoretical analysis and also propose a probabilistic approach to prove the outage probability can be effectively reduced to the minimal.

\end{abstract}
\IEEEpeerreviewmaketitle

\section{Introduction}
The pervasive use of wireless networks and the miniaturization have lead to the existence of WBANs. A person wears low power and cost sensor devices forms a star topology WBAN coordinated by single coordinator C. These sensors may be implanted inside or attached on human body. They can be used in various applications such as health monitoring, ubiquitous healthcare, sports and military. WBANs mainly monitor physical activities and capture vital signs as glucose percentage in blood, heart beats, respiration, body temperature and/or can record electrocardiography (ECG) \cite{key14,key15}.

Recently, the IEEE 802.15.6 working group has defined new PHY and MAC layer proposals for WBANs. The standard requires the whole system to maintain proper function when up to 10 WBANs are co-located within a transmission range of 3 meters \cite{key26}. It has also adopted the two-hop communication scheme in the standard. Thus, adopting relay transmission is a very promising solution for co-channel interference reduction, energy efficiency and high reliable communications \cite{key21,key7,key22}.

The unpredictable nature of WBANs and the high mobility make the coordination very hard. Due to the broadcast nature in WBANs, the nodes concurrently transmitting suffer from co-channel interference as acive periods can overlap. 

Co-channel interference problem motivates for the stringent requirements of interference avoidance schemes and protocols for reliable and energy efficient operation of WBANs. On the other hand, due to the constrained nature of WBANs (in terms of energy, size and cost), advanced antenna techniques can not be used for interference avoidance as well as power control mechanisms used in cellular networks are not applicable to WBANs \cite{key21,key22}. 

However, in this work, we focus our attention on problems related to co-channel interference and energy savings of a single WBAN. Thus, novel methods and schemes are required for intra-WBAN interference avoidance/mitigation. 

The rest of the paper is organized as follows. Section \rom{2} shows the works related to interference mitigation/avoidance techniques in WBANs. Section \rom{3} describes the system model and presents the proposed interference avoidance algorithm (IAA). Section \rom{4} explains the proposed FTDMA scheme. Section \rom{5} shows the theoretical analysis of our proposal. Section \rom{6} presents and explains the experimental results. The conclusions and future works are drawn in section \rom{7}.

\section{Related Works}
Recent studies show multi-hop schemes have a lower power consumption in comparison to one-hop scheme. However, using relays reduces the WBAN interference and consequently the power consumption. Authors of \cite{key7} propose a single-relay cooperative scheme where the best relay is selected in a distributed fashion. Also, authors of \cite{key17} propose a prediction-based dynamic relay transmission scheme through which the problem of "when to relay" and "who to relay" are decided in an optimal way. The interference problem among multiple co-located WBANs is investigated in \cite{key7}. The authors show cooperative two relay communication with opportunistic relaying significantly mitigates WBAN interference.

Authors of \cite{key22} investigate the problem of coexistence of multiple non coordinated WBANs. This study provides better co-channel interference mitigation. However, more recent works conducted in \cite{key9} propose a scheme for joint two-hop relay-assisted cooperative communication integrated with transmit power control. This scheme can reduce co-channel interference and extend the lifetime. 

On the other hand, other works prove that TDMA scheme is an attractive solution to avoid interference within an intra-WBAN. Authors of \cite{key6} enables two or three coexisting WBANs to agree on a common TDMA schedule to reduce the interference. The work in \cite{key3} adopts a TDMA polling-based scheme for traffic coordination inside a WBAN and a carrier sensing (CS) mechanism to deal with inter-WBAN interference.

Other research focuses on the performance at the coordinator that calculates SINR periodically. This calculation enables C to command its nodes to select appropriate interferece mitigation scheme \cite{key1}. Other studies of \cite{key6} analyze the performance of a reference WBAN. They evaluate the performance in terms of bit error rate, throughput and lifetime which have been improved by adoption of an optimized time hopping code assignment strategy. Works in \cite{key5} consider a WBAN where coordinator periodically queries sensors to transmit data. The network adopts the CSMA/CA and the nodes adopt link adaptation to select the modulation scheme according to the experienced channel quality. 

The research work of \cite{key23} solves the problem of inter-WBAN scheduling and interference by the adoption of a QoS based MAC preemptive priority scheduling approach. Whilst, researchers of \cite{key20} proposes a distributed interference detection and mitigation scheme through using adaptive channel hopping. Whereas, research works of \cite{key16} proposes a dynamic resource allocation scheme for interference avoidance among multiple coexisting WBANS through using orthogonal sub-channels for high interfering nodes.

Since TDMA is the most widely used protocol inside WBANs. Most of the recent works do not address problems related to interference minimization and energy maximization inside a WBAN. However, this protocol is not suitable for some applications with high transmission frequency and topology size of more then 12 body sensors. In this paper, we propose a distributed scheme (IAA) for interference avoidance inside a WBAN. The proposed scheme enables low interfering nodes to use base channel for transmitting to relays. Whilst, high interfering nodes extend their contention window and probably switch to another reserved channel.
\section{System Model}
We consider a fixed topology of a single WBAN consisting of a fixed set of N nodes and C. We adopt two-hop communication scheme and consider a beacon-enabled slotted CSMA/CA between sources and relays and a flexible TDMA between relays and C. However, each node (source or relay) can operate on either of base or reserved channels. Furthermore, two possible options [with- or without-] CW extension can be used with base channel depending on the interference level.
\subsection{Model Definitions}
We denote signal to interference and noise ratio (SINR) by $\delta$ and SINR threshold by $\delta_{Thr}$. Where $\delta$ is computed at a node-of-interest by the following equation \ref{sinr}:
\begin{equation}\label{sinr}
\delta =\cfrac{P_{RX}}{\sum_{i=1}^{N} I_i + N_0}
\end{equation}
 Where, $P_{RX}$ is the desired received power, $I_i$ is the received power from undesired transmitter i and $N_0$ is the noise. In addition, we also define the following sets: 
 \begin{itemize}
\item P: set of all sources that have sensed data to transmit in the contention access period (CAP) of the current frame
\item S: set of all sources, where, each source has $\delta$ $\geq$ $\delta_{Thr}$, (S $\subset$ P), i.e, there is no interference:
\begin{equation}
S = \{s_{i} \in S, \mid (\delta_{s_{i}} \geq \delta_{Thr}), \forall i\}
\end{equation}
\item IS: set of all interfering sources where, each source has $\delta$ $<$ $\delta_{Thr}$, (IS $\subset$ P), i.e, there is interference:
\begin{equation}
IS = \{s_{i} \in IS, \mid (\delta_{s_{i}} < \delta_{Thr}), \forall i\}
\end{equation}
\item TxR: set of all relays transmit successfully their messages to C in the current frame
\item $K_{i}$: set of all orthogonal channels reserved for relay $r_{i}$
\begin{equation}
K_{i} = \{ K_{i} \mid K_{i} \cap K_{j} = \varnothing,  \forall i \neq  j\}
\end{equation}
\item G: set of all orthogonal channels  reserved for all relays
\begin{equation}
 K_{i} \subset G \iff K_{i} \cap K_{j} = \varnothing, \forall i \neq j
\end{equation}
\end{itemize}
\subsection{Proposed Interference Avoidance Algorithm Description}
Each source $s_{i} \in S$ whose $\delta1_{s_{i}}$ $\geq$ $\delta_{Thr}$ (i.e. there is no interference experienced) uses the base channel to communicate directly to the relays \textbf{(case 1)}. Otherwise, source $s_{i} \in IS$ (i.e. there is an interference experienced) extends the CW (doubles its backoff) to avoid the current interference. Afterwards, any source has already finished its CW extension retries sensing again the base channel.
\IncMargin{1em}
\begin{algorithm}
\footnotesize
\SetKwData{Left}{left}\SetKwData{This}{this}\SetKwData{Up}{up}
\SetKwFunction{Union}{Union}\SetKwFunction{FindCompress}{FindCompress}
\SetKwInOut{Input}{input}\SetKwInOut{Output}{output}
\Input{P, $\delta_{Thr}$, $q_{Thr}$}

 \For{$i$ $\leftarrow$ $1$ $\KwTo$ $sizeof(P)$}
 {
 
  \If{($\delta1{s_{i}}$ $\geq$ $\delta_{Thr}$)}
    {

              $s_{i}$ $\in$ $S$ $sendsMessageOn$ baseChannel in CAP-1A;
     
    }
    
    \Else
    {
    
      $\therefore$ $s_{i}$ is an interfering source $\Leftrightarrow$ $s_{i}$ $\in$ $IS$;
       
       $s_{i}$ $doublesCW$ \& waits until CAP-1A finishes;
                           
         \If {$CW_{s_{i}}$ isOver}
            {
            
                  \If {$\delta2{s_{i}}$ $\geq$ $\delta_{Thr}$}
                  {
                  
                      $s_{i}$ $\in$ $IS$ $sendsMessageOn$ baseChannel in CAP-1B;
                  
                  }
                  
                 \Else
                  {
                  
                      $\therefore$ $s_{i}$ is an interfering source again;
                  
                      $s_{i}$ $switchesTo$ reservedChannel \& waits until CAP-1B finishes;

                  \If {$\delta3{s_{i}}$ $\geq$ $\delta_{Thr}$}
                  {
                  
                      $s_{i}$ $sendsMessageOn$ reservedChannel in CAP-2;
                  
                  }
                  
                  \Else
                  {
                      
                      q = 0;
                      
                      \For{$m$ $\leftarrow$ $1$ $\KwTo$ maxRetries}
                       {

                        \If {$\delta3{s_{i}}$ $<$ $\delta_{Thr}$}
                          {
                            q = q + 1;
                            
                          }
                          
                         \If {$q$ $>$ $q_{Thr}$}
                          {
                          
                             $s_{i}$ $switchesTo$ baseChannel;
                             
                             break;
                        
                          }
                            
                       }
                  
                  }
                  
                  }
                  
            }
               
    }
    
 }  
\caption{IAA - Sources Actions}
\label{algo_sources}
\end{algorithm}
\DecMargin{1em}

If it succeeds (finds $\delta2_{s_{i}}$ $\geq$ $\delta_{Thr}$), it transmits its message to relays using the base channel (\textbf{case 2}). Otherwise, source $s_{i} \in IS$ experiences an interference again and switches to reserved channel \textbf{(case 3)} through which it starts again a new contention (meausres $\delta3$). The following summarizes the different aforementioned cases.
\begin{itemize}
  \item {Case 1: $\forall$ $s_{i}$ $\in$ $S$ \&  $\delta1_{s_{i}}$ $\geq$ $\delta_{Thr}$, $s_{i}$ uses base channel}
  \item {Case 2: $\forall$ $s_{i}$ $\in$ $IS$ \& $\delta2_{s_{i}}$ $\geq$ $\delta_{Thr}$, $s_{i}$ uses base channel with CW extension mechanism}
  \item {Case 3: $\forall$ $s_{i}$ $\in$ $IS$ \& $\delta2_{s_{i}}$ $<$ $\delta_{Thr}$, $s_{i}$ uses reserved channel}
 \end{itemize}
The Algo. \ref{algo_sources} above shows the pseudocode of the IAA part runs at the sources.
\subsection{Proposed Superframe Structure}
To support our proposed scheme, a new superframe structure is composed of two main parts, a CAP part and TDMA part as shown in Fig. \ref{frame}. The CAP part is composed of CAP-1 and CAP-2 sub-parts. Whereas, the TDMA part is composed of fixed and flexible TDMA sub-parts. CAP-1 is also composed of CAP-1A (covers \textbf{cases 1}) and CAP-1B (covers \textbf{cases 2}) sub-parts. For thoroughly explanation, all the sources' transmissions done on base channel (\textbf{cases 1} and \textbf{2}) must complete just before the end of CAP-1. Whereas, all sources' transmissions done on reserved channel (\textbf{case 3}) must start just after the end of CAP-1 and complete just before the end of the second period CAP-2. On the other hand, after all the sources' transmissions to the relays complete in both CAP-1 and CAP-2. The TDMA part starts through which some set of relays commence transmitting their pending messages to C using their corresponding slots of the TDMA schedule. Regarding the fixed part of the frame, a pre-defined fixed number of time slots is assigned to a specific set of relays. Based on the network interference level in the current frame, C estimates a flexible number of time slots for the next frame that will be used by an unexpected set of relays.

\subsection{Source to Relay Communication }
\begin{figure}[ht]
 \centering
   \includegraphics[width=0.4\textwidth]{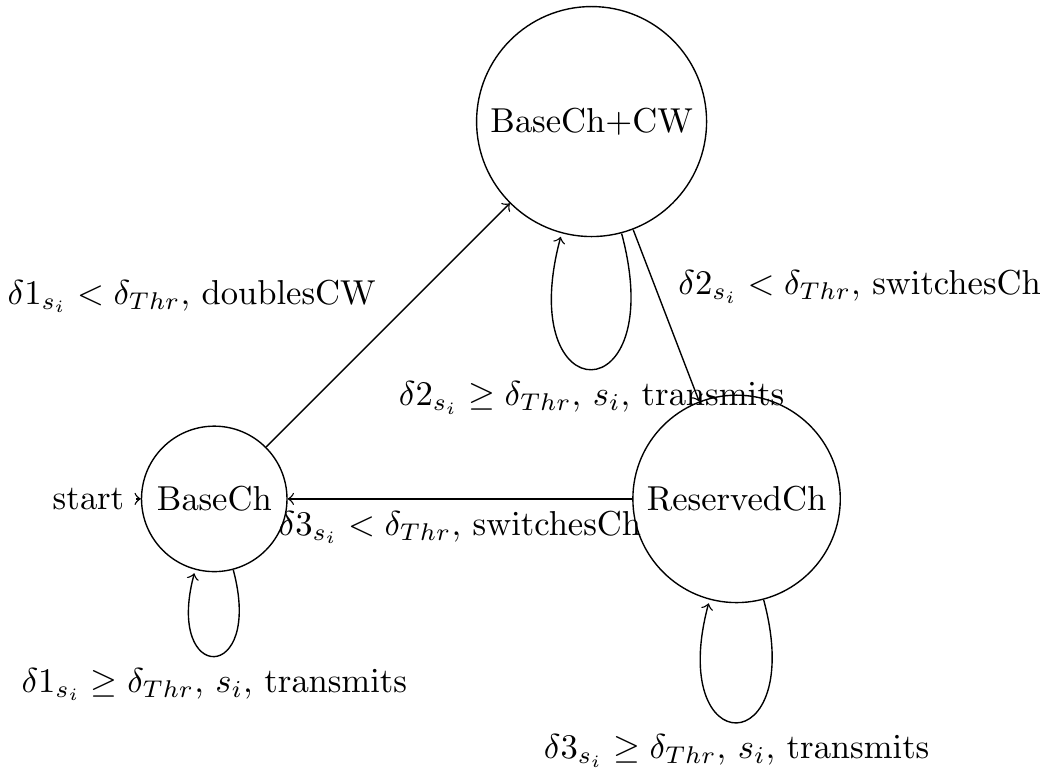}
     \caption{Actions taken by a source at any CAP period }
     \label{fsm}
\end{figure}

The communication between sources and relays is achieved through three successive periods. During the period CAP-1A, each source $s_{i}$ and whose $\delta1_{s_{i}}$ $\geq$ $\delta_{Thr}$ uses the base channel to transmit its message to the relays \textbf{(case 1)}. Otherwise, $s_{i}$ is considered interfering, if it is so, it extends its CW to avoid the interference. It waits until CAP-1A finishes. Afterwards, it retries and if finds its $\delta2_{s_{i}}$ $\geq$ $\delta_{Thr}$, it uses the base channel again through which it transmits its message to the relays during the the CAP-1B \textbf{(case 2)}. Otherwise, if it finds its $\delta2_{s_{i}}$ $<$ $\delta_{Thr}$, it switches to the reserved channel through which it starts a new contention commencing just after the end of CAP-1B \textbf{(case 3)}. However, Fig. \ref{fsm} shows  all the different possible actions taken by a source at any CAP period.
\subsection{Relay Actions and Channel Synchronization}
To let the sources have already switched to reserved channel transmitting their messages correctly. Our proposal ensures that there are some set of relays ready to receive these transmissions. Initially, all the relays (R) listen on base channel. Each relay $r_{i}$ $\in$ R measures periodically $\delta_{r_{i}}$ in a pre-defined segment of each CAP, if it finds $\delta_{r_{i}}$ $\geq$ $\delta_{Thr}$, then $r_{i}$ can receive on base channel. Otherwise, if it finds $\delta_{r_{i}}$ $<$ $\delta_{Thr}$, i.e, $r_{i}$ experiences an interference, it then switches to the reserved channel where it starts listening again. Whenever a relay encounters a collision, it immediately transmits a jam signal to inform the transmitting sources to stop transmitting and waits a while (simple backoff) and then retries \cite{key27}. According to the relays' actions aforementioned, the same process takes place at both base and reserved channels. Afterwards, when all receptions at relays are complete, each relay waits TDMA to commence transmitting to C. The following Algo. \ref{algo_relays} shows the pseudocode of the IAA part runs at the relays.
\begin{figure}[ht]
 \centering
   \includegraphics[width=0.4\textwidth]{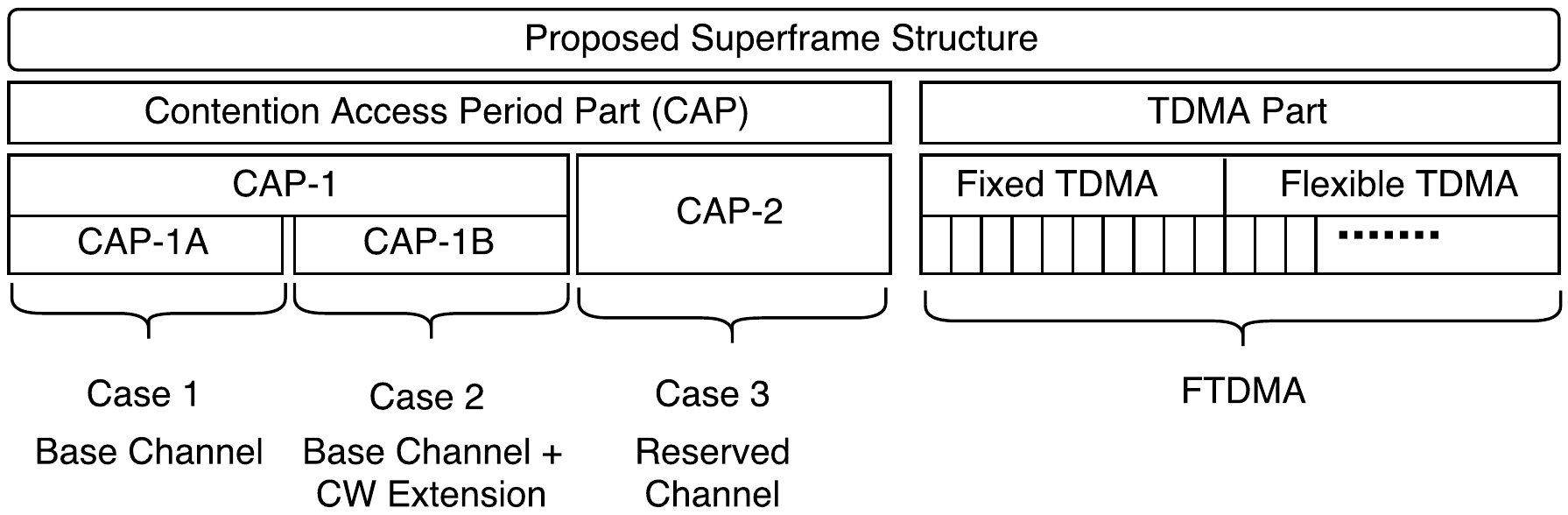}
     \caption{Proposed Superframe Structure}
     \label{frame}
\end{figure} 
\subsection{Relays to Coordinator Communication}
After all the transmissions are complete, all the sources and relays switch back to base channel. However, according to the aforementioned proposed FTDMA scheme, C forms the fixed part of the next frame consisting of TS time slots, where, \textbf{TS = B + Re + BW}, \textbf{B} is the number of slots assigned for nodes relaying data on the behalf of the sources that use the base channel \textbf{(case 1)}. \textbf{Re} is the number of slots assigned for nodes relaying data on the behalf of the sources that use the reserved channel \textbf{(case 3)} and \textbf{BW} is the number of slots assigned for nodes relaying data on the behalf of the sources that use the base channel with CW extension \textbf{(case 2)}. Then, depending on the interference level experienced in the previous superframe, C specifies and adds the number of free slots of the flexible part to the fixed part to form the whole superframe. The following Algo. \ref{cactions} shows the pseudocode of the IAA part runs at the coordinator.
\IncMargin{1em}
\begin{algorithm}
\footnotesize
\SetKwData{Left}{left}\SetKwData{This}{this}\SetKwData{Up}{up}
\SetKwFunction{Union}{Union}\SetKwFunction{FindCompress}{FindCompress}
\SetKwInOut{Input}{input}\SetKwInOut{Output}{output}
\Input{TxR, Interference-level $IL_{k}$}

     C Broadcasts Beacon $b_{k}$
     
     m = 0;
            
    \For{$i\leftarrow 1$ $\KwTo$  $sizeof(TxR)$}
       {
     
          \If {C Acknowledges $rs_{i}$} 
            {
          
              C includes $ID_{rs_{i}}$ in fixed TDMA part of Beacon $b_{k + 1}$
              
              m = m + 1;
          
             }
        
       }

       C forms fixed TDMA part of m slots
       
       C forms flexible TDMA part of n slots based on $IL_{k}$
       
       C forms next Beacon $b_{k + 1}$ of (p = m + n) slots

\caption{IAA - Coordinator Actions}
\label{cactions}
\end{algorithm}
\DecMargin{1em}
\section{Proposed Flexible TDMA Scheme (FTDMA)} 
A node is considered active if C has received at least one message during the previous three frames. If C has received from m active nodes in the current frame, then, it allocates p (where p > m) slots in TDMA part of the next frame. The first m slots are allocated to the currently active nodes (fixed part) and the rest (n = p - m) slots are reserved to the newly incoming nodes (flexible part). See Fig. \ref{frame}.
 
A node first listens to the beacon, if it finds its ID in the fixed TDMA part, it transmits its message in one of the m slots. However, if it does not find its ID, it then randomly selects one of the n empty slots (flexible TDMA part) and transmits its meassage in that slot. If the message is successfully sent to C, C will allocate a slot for the node in the next frame. Otherwise, C will not include its ID in the next frame. In such cases, a node keeps trying different empty slots randomly in every frame until a timeslot is assigned to it. Algo. \ref{cactions} shows how C assigns slots to nodes in the next superframe.

\section{Theoretical Analysis of Proposed IAA Scheme}
Outage probability is a metric for the channel that states according to the variable $\delta$ at the received end, what is the probability that a capacity is not supported due to variable $\delta$. In other words, outage probability denoted by (OP) at given $\delta_{Thr}$ is defined as the probability of $\delta$ value being larger than threshold $\delta_{Thr}$. 
\begin{equation}
OP = Pr\left(\delta > \delta_{Thr}\right)
\end{equation}
We denote by $P_{out}$ the probability that the total interference at time instant i is being larger than $\delta_{Thr}$ at a given source s of the WBAN. We denote by $\delta_{j}$ the received  $\delta$ from sensor j at sensor s in WBAN. Then, we calculate this probability by the following formula:
\begin{equation}
P_{out} = \left(\displaystyle\sum_{j=1}^{N-1} \delta_{j} > \delta_{Thr} \right)
\end{equation}
We present a probabilistic approach which we prove analytically it lowers the outage probability. Any sensor s whose received $\delta$ is lower than a given threshold, it doubles its contention window iff $\delta_{j}$ $<$ $\delta_{Thr}$. So, sensor s extends its CW with certain probability which equals $\frac{\delta_{j}}{\delta_{Thr}}$. Thus, at time instant i, we can calculate the average interference level at source s using the proposed probabilistic approach as follows:
\begin{equation}
\delta_{i} = \displaystyle\sum_{j=1}^{N-1} \delta_{j} \left(1 - \frac{\delta_{j}}{\delta_{Thr}} \right)
\end{equation}
Based on the probabilistic approach and the proposed scheme, any sensor with probability $\frac{\delta_{j}}{\delta_{Thr}}$ doubles its contention window. If the source is in contention window case, it then probably (depending on $\delta$) switches to the reserved channel with probability of $\left(\frac{\delta_{j}}{\delta_{Thr}}\right)^2$. 

\textbf{Lemma 1:} We denote by $P_{Pr}$ and $P_{out}$ the outage probability of probabilistic approach and the outage probability of the original scheme respectively. Then, $P_{Pr}$ < $P_{out}$, i.e. the probabilistic approach has better $\delta$ than that of the original scheme.

\textbf{Proof:} Based on outage probability definition, we have:
\begin{equation} 
P_{Pr} = p \left(\displaystyle\sum_{i=1}^{N-1} \delta_{i} \left(1 - \left(\frac{\delta_{i}}{\delta_{Thr}} +  \left(\frac{\delta_{i}}{\delta_{Thr}}\right)^2 \right)\right) > \delta_{Thr}\right) 
\end{equation}
\begin{equation} 
= p \left(\displaystyle\sum_{i=1}^{N-1} \delta_{i} > \delta_{Thr} + \displaystyle\sum_{i=1}^{N-1}\frac{\delta_{i}^2}{\delta_{Thr}} + \displaystyle\sum_{i=1}^{N-1}\frac{\delta_{i}^3}{\delta_{Thr}^2} \right)
\end{equation}  
\begin{equation}             
< p  \left( \displaystyle\sum_{j=1}^{N-1} \delta_{j} >  \delta_{Thr} \right) = P_{out}  
\end{equation}

Where $\left(\frac{\delta_{j}}{\delta_{Thr}} + \left(\frac{\delta_{j}}{\delta_{Thr}}\right)^2\right)$ denotes the probability of the source is being in case 2 \textbf{or} the probability of the source is being in case 3. The last line of $P_{Pr}$ is based on the fact that the CDF is an increasing function of its argument. We define $P_{pr, I, i}$ as the probabilistic approach deployment probability that a sensor node of WBAN doubles its contention window is Then:
\begin{equation} 
P_{prob, I, i} = P(\delta_{i} > \delta_{Thr}) + P(\delta_{i} < \delta_{Thr})\frac{\delta_{i}}{\delta_{Thr}},   
\end{equation}
which is greater than $P_{i}$ = $P(\delta_{avg} > \delta_{Thr})$.
\IncMargin{1em}
\begin{algorithm}
\footnotesize
\SetKwData{Left}{left}\SetKwData{This}{this}\SetKwData{Up}{up}
\SetKwFunction{Union}{Union}\SetKwFunction{FindCompress}{FindCompress}
\SetKwInOut{Input}{input}\SetKwInOut{Output}{output}
\Input{TxR, $\delta_{Thr}$}

       \For{$k$ $\leftarrow$ $1$ $\KwTo$ sizeof($TxR$)}
       {
        
          $r_{k}$ listensOn baseChannel;
        
                    \If{$\delta1{r_{k}}$  $\geq$ \textbf{\textit{$\delta_{Thr}$}}}
                    {
                              
                         $r_{k}$ receivesOn baseChannel;
                        
                   }
                        
                   \Else
                   {
                        
                            $r_{k}$ switches$\&$ListensOn reservedChannel;
                        
                           \If{$\delta2{r_{k}}$  $\geq$ \textbf{\textit{$\delta_{Thr}$}}}
                            {
                            
                            $r_{k}$ receivesOn reservedChannel;
                            
                            }

                           \Else
                             {
                   
                               $r_{k}$ switchesTo baseChannel after $maxRetries$;
                   
                             }
                        
                    }
                        
       }
           
\caption{IAA - Relays Actions}
\label{algo_relays}
\end{algorithm}
\DecMargin{1em}

\section{Experimental Results}
\subsection{Simulation Environment and Setup}
 We have considered a static WBAN consisting of N = 12 source nodes and a set of relay nodes R = 4 located in an area of 2$\times$2 $m^{2}$. All the nodes operate in a half-duplex mode and use the same transmission power at 0 dBm. In this simulation, we focus our attention on the performance of three important metrics; minimum SINR, outage probability and WBAN energy residue. For brevity, the rest of simulation parameters are listed in table \ref{tab:title} above.
\begin{table}
\centering
\normalsize
 \begin{tabular}{ |p{3.5cm}|p{3.5cm}|  }
\hline
Simulation time & 50 minutes \\\hline
Noise floor & -100 dBm \\\hline
Data rate & 250 kbps  \\\hline
Packet size & 12 bytes \\\hline
Frequency & 2.4 GHz  \\\hline
Path loss exponent($\alpha$) &  4.22 \\\hline
\end{tabular}
 \caption {Simulation Parameters} \label{tab:title}
\end{table}
\subsection{WBAN Minimum SINR}
\begin{figure}[ht]
 \centering
   \includegraphics[width=0.4\textwidth]{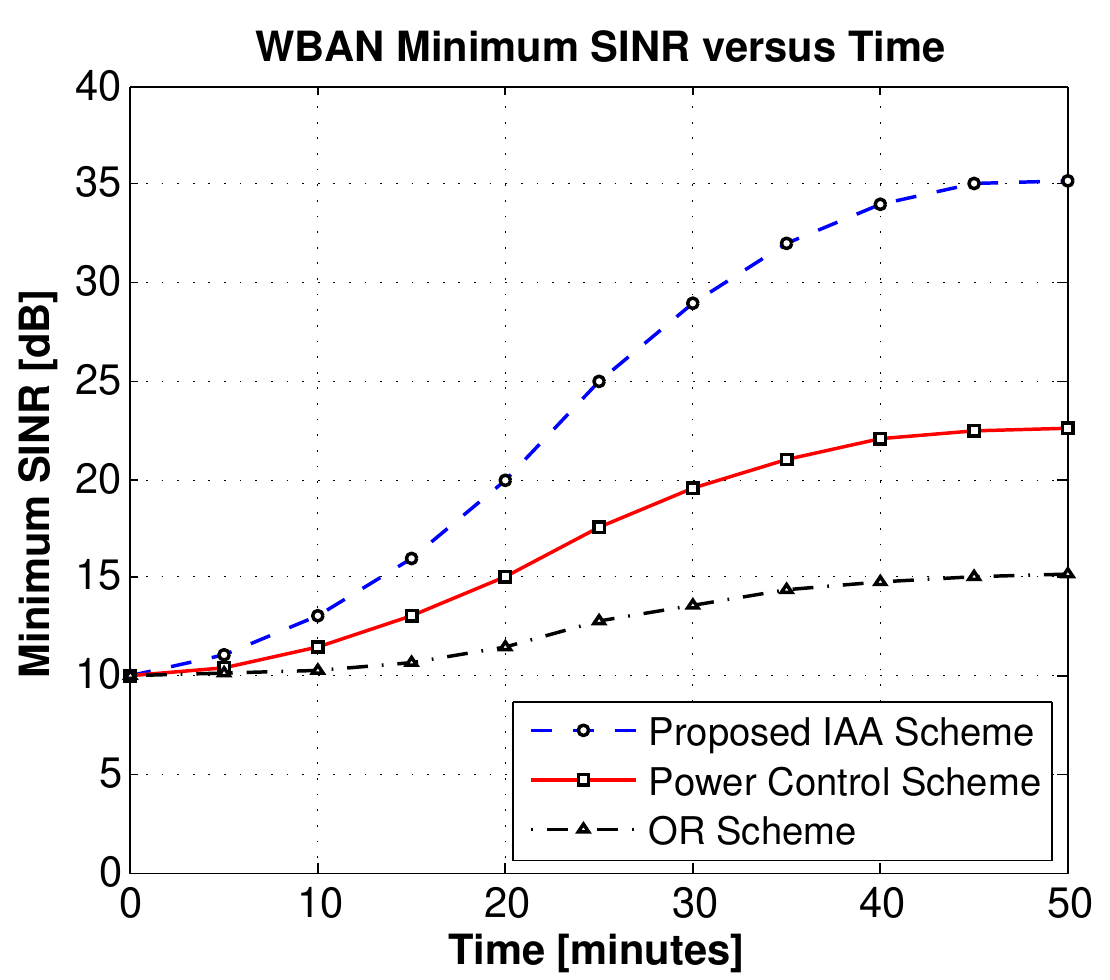}
     \caption{WBAN Minimum SINR versus Time of proposed IAA scheme compared to that of power control (PC) and opportunistic relaying (OR) schemes}
     \label{iaasinr}
\end{figure}

We have chosen the most common metric SINR to evaluate the interference level and show how better our proposed IAA scheme mitigates the interference than other schemes. The minimum SINR ($\delta_{min}$) of the WBAN versus time for IAA scheme is compared to the cooperative communication integrated with transmit power control (PC) \cite{key9} and CSMA/CA with opportunistic relaying (OR) \cite{key7,key27} schemes. As can be clearly seen in Fig. \ref{iaasinr}, IAA scheme achieves a higher $\delta_{min}$ (12dB improvement) than PC and OR schemes. PC provides a higher $\delta_{min}$ (7dB improvement) than OR scheme since the former adopts the power control mechanism where the nodes (sources and relays) dynamically adjust their power level. Accordingly, controlling power reduces the interference at other nodes of WBAN and hence improving their $\delta_{min}$. When some nodes experience high interference, they have to wait the channel to become free and then retry. However, instead of waiting, in the proposed IAA scheme, these high interfering nodes extend the CW and probably switch to reserved channel for avoiding interference and so their corresponding $\delta_{min}$s are increased. Furthermore, only IAA use flexible TDMA to communicate to C for avoiding the interference. As a result, our proposed IAA scheme avoids the WBAN interference through increasing the minimum SINR.
\subsection{Outage Probability}
\begin{figure}[ht]
 \centering
   \includegraphics[width=0.4\textwidth]{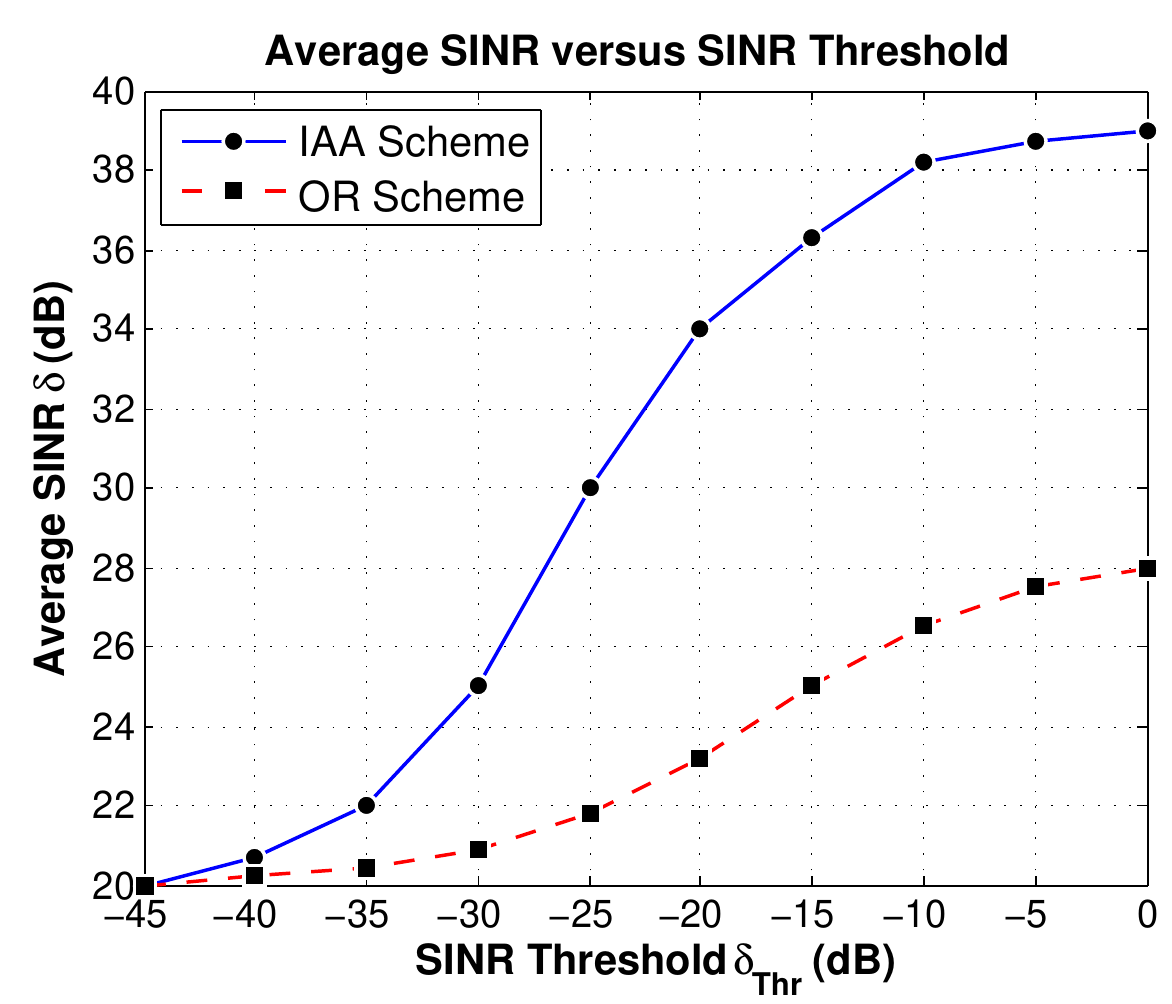}
     \caption{Average SINR versus SINR Interference Threshold of the proposed IAA scheme compared to that of opportunistic relaying (OR) scheme}
     \label{op}
\end{figure}

We have evaluated the average SINR $\delta$ of the WBAN versus the SINR threshold $\delta_{Thr}$ in Fig. \ref{op}, where a higher $\delta$ which corresponds to a lower outage probability is obtained when the interference threshold is increased. This figure compares the average $\delta$ for the proposed IAA scheme and that for the OR scheme. As can be clearly seen in this figure, the proposed IAA scheme achieves a higher average $\delta$ for all SINR interference thresholds greater than -45 dB which is quite efficient in the case of WBANs. Thus, increasing SINR interference thresholds puts more sensors in the interference set. These sensors will use CW extension mechanism and probably the reserved channel to avoid the interference and hence improving their SINRs. In addition, it is evident to see, when the interference threshold reaches -45 dB, the proposed IAA scheme approaches the OR scheme in terms of average $\delta$. However, it is important to note that the interference threshold can be adaptively selected according to the interference level. 
\subsection{WBAN Energy Residue}
\begin{figure}[[ht]
 \centering
   \includegraphics[width=0.4\textwidth]{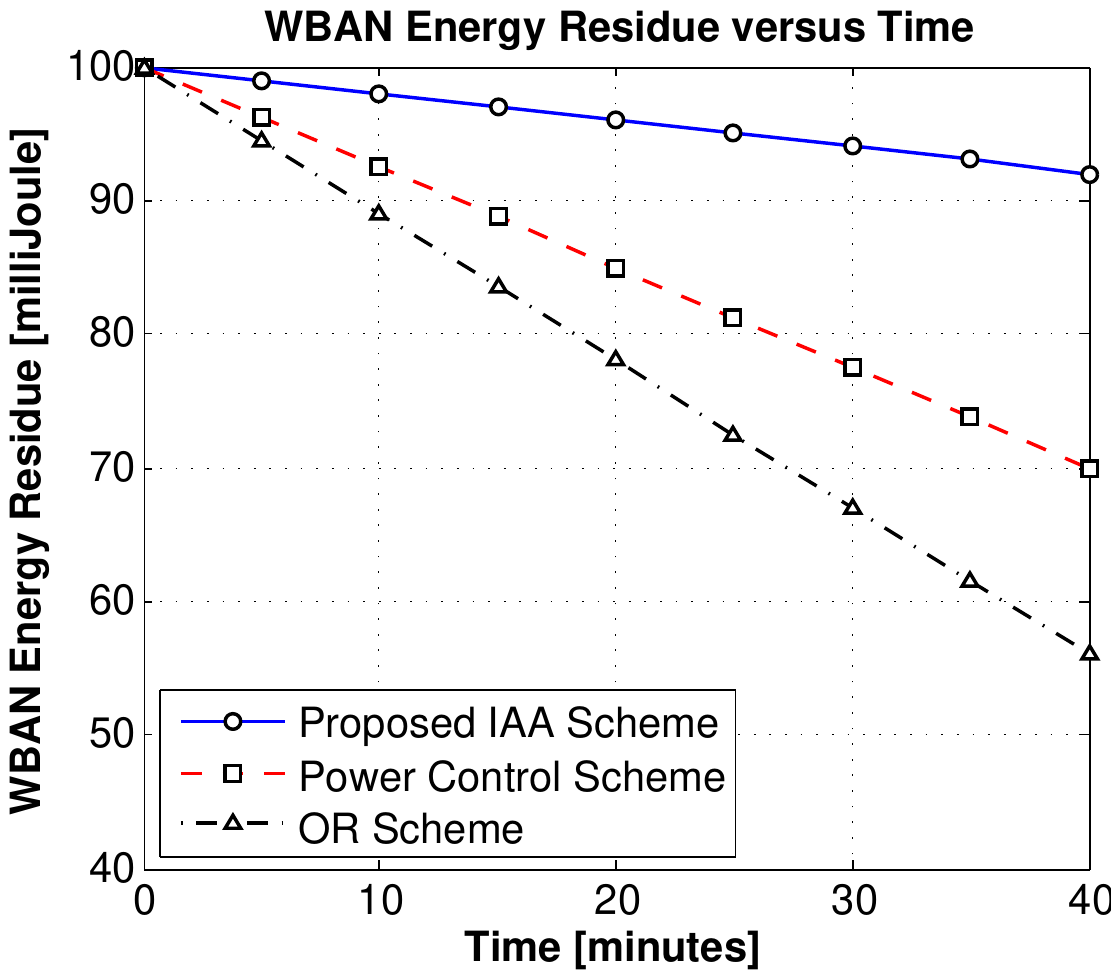}
     \caption{WBAN Energy Residue versus Time of the proposed IAA scheme compared to that of power control (PC) and opportunistic relaying (OR) schemes}
     \label{fig:energy}
\end{figure}

In order to evaluate our proposed IAA scheme how better extends the energy lifetime of the WBAN than other schemes. We have chosen energy residue which is linearly related to the most stringent factor (energy) in WBANs. We define WBAN energy residue (ER) at time t as the sum of remaining energies in the battery of each sensor of the WBAN. ER versus time for three different schemes is compared and shown in Fig. \ref{fig:energy}. As can be clearly seen in this figure, ER of IAA scheme outperforms and is always higher than ER of OR and PC schemes. ER of IAA decreases slightly whilst in the other schemes decreases sharply. This ensures a longer WBAN energy lifetime. However, with the proposed IAA scheme, the packet collisions and retransmissions are also minimized because all the interfering nodes dynamically use the CW extension mechanism and probably the channel switching technique. In PC scheme, a power control mechanism adjusts dynamically the power level at sources and relays and hence reducing the energy consumption in the whole WBAN. On the other hand, the absence of power control from OR scheme increases the energy consumption at the sources and relays and thus shortening the energy lifetime of the whole WBAN. As a result, our proposed IAA scheme outperforms other schemes and better extends the WBAN's energy lifetime. Furthermore, adopting a flexible TDMA (PC and OR do not adopt) in our proposed scheme to avoid interference decreases also the whole WBAN energy consumption. 

\section{Conclusion}
In this paper, we propose a distributed cooperative multi-hop scheme (IAA) adopts contention window extension mechanism integrated with channel switching technique to avoid co-channel interference of a single WBAN. We also propose a flexible TDMA between relays and coordinator. Our proposal aims to avoid co-channel interference and extends the WBAN energy lifetime. Furthermore, our proposed IAA scheme has been evaluated by simulation and compared with other schemes showing 12dB improvement in the minimum SINR and significantly extends the WBAN energy lifetime. Additionally, theoretical analysis validated our approach. Thus, we propose a probabilistic approach and prove the outage probability is reduced to the minimal. As a future work, it is worth to adapt our proposal for the dynamic topology taking into account the body movement in an environment of multiple coexisitng WBANs. 


\begin{thebibliography}{16}
\bibitem{key1}
  Yang, Wen-Bin and Sayrafian-Pour, Kamran, Interference Mitigation Using Adaptive Schemes in Body Area Networks, International Journal of Wireless Information Networks, pages 193-200, 2012
  
\bibitem{key3}
  Chen, G. Chen, W. and Shen, S., 2L-MAC: A MAC Protocol with Two-Layer Interference Mitigation in Wireless Body Area Networks for Medical Applications, IEEE International Conference on Communications (ICC), pages 3523-3528, 2014
  
\bibitem{key4}
Xuan Wang and Lin Cai, IEEE Global Telecommunications Conference (GLOBECOM), 2011, Interference Analysis of Co-Existing Wireless Body Area Networks,
2011

\bibitem{key5}
Martelli, F. and Verdone, R. and Buratti, C., Link Adaptation in IEEE 802.15.4-based Wireless Body Area Networks, Personal, Indoor and Mobile Radio Communications Workshops (PIMRC Workshops), 2010 IEEE 21st International Symposium, pages 117-121,2010

\bibitem{key6}
Mahapatro, J. and Misra, S. and Manjunatha, M. and Islam, N., Interference mitigation between WBAN equipped patients, Ninth International Conference on Wireless and Optical Communications Networks (WOCN),pages 1-5,2012

\bibitem{key7}
  Dong, Jie and Smith, David, Opportunistic relaying in wireless body area networks: Coexistence performance, IEEE International Conference on Communications (ICC), pages 5613-5618, 2013

\bibitem{key8}
Vakil, S. and Ben Liang, Relaying in Wireless Sensor Networks with Interference Mitigation,IEEE Global Telecommunications Conference (GLOBECOM, 2006

\bibitem{key9}
  Dong, Jie and Smith, David,Joint relay selection and transmit power control for wireless body area networks coexistence, IEEE International Conference on Communications (ICC), pages 5676-5681, 2014
  
\bibitem{key10}
Braem, B. and Latre, B. and Moerman, I. and Blondia, C. and Reusens, E. and Joseph, W. and Martens, L. and Demeester, P., The Need for Cooperation and Relaying in Short-Range High Path Loss Sensor Networks,  International Conference on Sensor Technologies and Applications, SensorComm, 2007

\bibitem{key14}
 Javaid, Nadeem and Khan, NA and Shakir, M and Khan, MA and Bouk, Safdar Hussain and Khan, ZA, Ubiquitous healthcare in wireless body area networks-a survey,
 arXiv preprint arXiv:1303.2062, 2013

\bibitem{key15}
Movassaghi, Samaneh and Abolhasan, Mehran and Lipman, Justin and Smith, David and Jamalipour, Abbas, Wireless Body Area Networks: A Survey,
IEEE Communications Surveys Tutorials, pages 1658-1686, 2014
\bibitem{key16}
  Movassaghi, Samaneh and Abolhasan, Mehran and Smith, David, Smart spectrum allocation for interference mitigation in Wireless Body Area Networks,
  IEEE International Conference on Communications (ICC), pages 5688-5693, 2014
 
\bibitem{key17}
Feng, Hui and Liu, Bin and Yan, Zhisheng and Zhang, Chi and Chen, Chang Wen, Prediction-based dynamic relay transmission scheme for Wireless Body Area Networks, IEEE 24th International Symposium Personal Indoor and Mobile Radio Communications (PIMRC), pages 2539-2544,2013

\bibitem{key20}
Shipeng Liang and Yu Ge and Shengming Jiang and Hwee Pink Tan, A lightweight and robust interference mitigation scheme for wireless body sensor networks in realistic environments, IEEE Wireless Communications and Networking Conference (WCNC), pages 1697-1702, 2014

\bibitem{key21}
  Movassaghi, Samaneh and Abolhasan, Mehran and Smith, David, Interference Mitigation in WBANs: Challenges and Existing solutions, Workshop on Advances in Real-time Information Networks,2013

\bibitem{key22}
Jie Dong and Smith, D., Cooperative body-area-communications: Enhancing coexistence without coordination between networks, IEEE 23rd International Symposium on Personal Indoor and Mobile Radio Communications (PIMRC), pages 2269-2274, 2012

\bibitem{key23}
 Jamthe, Anagha and Mishra, Amitabh and Agrawal, Dharma P, Scheduling schemes for interference suppression in healthcare sensor networks, IEEE International Conference on Communications (ICC), pages 391-396, 2014


\bibitem{key26}
IEEE Standard for Local and metropolitan area networks - Part 15.6: Wireless Body Area Networks,pages 1-271,2012

\bibitem{key27}
  Dong, Jie and Smith, David, Coexistence and Interference Mitigation for Wireless Body Area Networks: Improvements using On-Body Opportunistic Relaying, 
  arXiv preprint arXiv:1305.6992, 2013

\end{thebibliography}

\end{document}